# Determinants of Hotels and Restaurants entrepreneurship: A study using GEM data

Antonio Rafael Ramos-Rodríguez [a,*], José Aurelio Medina-Garrido [a], José Ruiz-Navarro [b]

[a] Faculty of Social Science and Communication, University of Cádiz, Jerez de la Frontera, Cádiz, Spain
[b] Faculty of Economic and Management Sciences, University of Cádiz, Cádiz, Spain
* Corresponding author: rafael.ramos@uca.es (A.R. Ramos-Rodríguez).

This is the submitted version accepted for publication in the "International Journal of Hospitality Management". The final published version can be found at:

https://doi.org/10.1016/j.ijhm.2011.08.003

We acknowledge that Elsevier Science LTD holds the copyright of the final version of this work. Please, cite this paper in this way:

Ramos-Rodríguez, A. R., Medina-Garrido, J. A., & Ruiz-Navarro, J. (2012). Determinants of Hotels and Restaurants entrepreneurship: A study using GEM data. International Journal of Hospitality Management, 31(2), 579–587. https://doi.org/10.1016/j.ijhm.2011.08.003

**ABSTRACT**

The objective of this work is to assess the influence of certain factors on the likelihood of being a Hotels and Restaurants (H&R) entrepreneur. The factors evaluated are demographic and economic variables, variables related to perceptions of the environment and personal traits, and variables measuring the individual's intellectual and social capital. The work uses logistic regression techniques to analyze a sample of 33,711 individuals in the countries participating in the GEM project in 2008. The findings show that age, gender, income, perception of opportunities, fear of failure, entrepreneurial ability, knowing other entrepreneurs and being a business angel are explanatory factors of the probability of being an H&R entrepreneur.

**INTRODUCTION**

The academic literature has been paying increasing attention to the phenomenon of firm creation in recent years. There is now recognition that this has a positive impact on the generation of both wealth and employment in the country (Acs et al., 2005; Reynolds et al., 2003). In particular, tourism entrepreneurship is proving to be highly important for economic development (Ateljevic, 2009) in both industrialized and developing countries (Mottiar and Ryan, 2007). But despite this, few works have analyzed the determinants of firm creation in the tourism industry (Ateljevic and Page, 2009).

Tourism is a broad concept that means different things to different people. Because of its complexity and interdisciplinary nature there is no adequate theoretical framework to study it as a distinct economic and social phenomenon.

The tourism industry contains a wide variety of different businesses involving the creation of tourism packages (e.g., travel wholesalers), sale and intermediation (e.g., travel agencies), transport of passengers, Hotels and Restaurants, and so on. This heterogeneity implies different entrepreneurial behaviors (Szivas, 2001), and different determinants of the probability of being an entrepreneur in this industry, so analysis of the tourism entrepreneur requires the selection of a more specific unit of analysis. Thus the current work focuses only on hospitality entrepreneurs in the Hotels and Restaurants (H&R) segments.

There are not many researchs that have studied entrepreneurship in these two segments of the hospitality industry. Berger and Bronson (1981) studied 50 restaurant entrepreneurs and discovered, in addition to the "craftsmen" and "opportunistic" entrepreneurs, a third category of "humanistic entrepreneurs". This group is characterized by a high desire to interact with people and a genuine concern for employees. Chell and Pittaway (1998) introduced the Critical Incident Technique to explains the method and shows how the method can be used to study entrepreneurial behaviour in the restaurant and café industry. This paper reported the initial results of a study of entrepreneurship in the Newcastle upon Tyne restaurant and cafe industry. This technique and the initial analysis of the data enabled them to reveal in a systematic way the heterogeneity of the small business owner population in the restaurant and café industry, and relate the information to business performance.

This work uses a concept of entrepreneur that has found some consensus in the literature: early-stage entrepreneur (Acs et al., 2005; Wagner, 2004; Davidsson and Honig, 2003). The GEM project defines the early-stage entrepreneur to be an entrepreneur that starts up a new business but has not paid salaries or wages for more than 3.5 years (Reynolds et al., 2005).

The concept of entrepreneur encompasses both the individual entrepreneur and the corporate entrepreneur (Sharma and Chrisman, 1999). The study of the entrepreneurial behavior of both types of entrepreneurs could be equally interesting. However, given that the variables to be studied, the research design and the results of both studies differ, this paper has focused only on the individual entrepreneur, leaving the analysis of the behavior of the corporate entrepreneur for later studies (Burgelman, 1983b, 1983a; Zahra and Covin, 1995; Zahra, 1991).The objective of the current work is to study the factors influencing individuals to take the decision to create an H&R business. In other words, the work aims to explore and explain individuals' propensity to create an H&R business.

The triggers of the intention or propensity to create a firm cannot, however, be studied using only objective variables (age, gender, income, etc.). Other variables have a role in this process, for example those to do with individuals' motivations, or their perceptions about their environment. Thus this work adds individuals' attitudes, perceptions and personal attributes to the analysis. This approach is completely new in this sector, and coherent with the one adopted by Arenius and Minniti (2005), who combine demographic and economic factors with perceptual variables referring to the entrepreneur.

The next section describes the most appropriate theoretical frameworks for the current work and formulates various hypotheses about how the propensity to be an H&R entrepreneur is affected by certain demographic, economic, perceptual, and social and intellectual capital factors. These hypotheses are based on some of the most important theories for the objectives of this research described in the second section. Section 3 describes the methodological aspects of the research: the characteristics of the sample analyzed, the variable measurement and the statistical model used. Section 4 reports the results obtained and discusses the outcome of the hypothesis tests. The final section offers the conclusions of this research and future research.

**THEORETICAL FRAMEWORK AND HYPOTHESES**

The diversity of disciplines studying the firm creation phenomenon (economics, business management, sociology, psychology, etc.), and the large number of paradigms adopted, are proof that none predominates and all suffer from some limitations. This means that the phenomenon should be studied from an eclectic perspective. Thus the current work combines various theoretical perspectives, considering their numerous dimensions, in order to be able to explain the determinants of individuals' propensity to create a firm.

Following Veciana (2007), the various theories on firm creation can be grouped into economic, psychological, socio-cultural and management theories. The economic theories explain the entrepreneurial function and firm creation on the basis of economic rationality. The psychological theories analyze how individuals' traits and characteristics affect the firm creation phenomenon. The socio-cultural theories assume that environmental factors condition the propensity to be an entrepreneur. Finally, the management theories consider that the decision to create a firm follows a rational process based on techniques, useful knowledge and practical models oriented to action.

The most appropriate of these four groups for the objectives of the current research are the psychological and socio-cultural theories. This is due to the perceptual nature of some of the determinant factors of entrepreneurial behavior. Although the environment surrounding the H&R entrepreneur can undoubtedly also be decisive, individuals clearly act in function of their perception of that environment. Psychological and socio-cultural factors intervene here, making this perception different from individual to individual. These factors include perceptions of the environment, personal traits, and intellectual and social capital.

The independent variables examined here for their effect on H&R entrepreneurship are classified in three groups: (1) demographic and economic factors; (2) variables related to perceptions of the environment and personal traits; and (3) intellectual and social capital factors. The demographic and economic factors help to complete the profile of the H&R entrepreneur. The second group contains variables describing individuals' perception of the environment (Arenius and Minniti, 2005) and personal traits (fear of failure). The third group consists of the variables confidence in one's skills, education level and whether the H&R entrepreneur knows any other entrepreneurs as part of their social network or as a business angel (Davidsson and Honig, 2003).

**Demographic and economic factors**

*Age*

Adopting a demographic perspective, some researchers have analyzed the relation between age and propensity to create a firm. Reynolds et al. (2003) and Blanchflower (2004) find that the entrepreneur tends to be young (between 25 and 34 years), although established entrepreneurs are normally older. Thus individuals' propensity to create a firm seems to be directly related to how young they are. In this respect, the first hypothesis is as follows:

Hypothesis 1: The older the individuals, the lower their propensity to create a firm in H&R sector.

Hypothesis 1 seeks to analyze whether entrepreneurial intention in this sector decreases with age. However, there are exceptional cases that depart from the norm. In this regard, it is worth highlighting cases such as Donald Trump or Robert Mondavi, who became entrepreneurs at an older age than normal. Moreover, the demographic evolution of the most developed countries and the aging of their populations are further fueling this phenomenon (Weber and Schaper, 2004; Minerd, 1999).

*Gender*

Men and women do not appear to have significant psychological differences (Langowitz and Minniti, 2005), although their entrepreneurial objectives and management styles do diverge (Brush, 1990, 1992). Nevertheless, from a demographic perspective empirical work has shown that men create more firms than women (Reynolds et al., 2003; Arenius and Minniti, 2005). Thus the current work examines whether the following hypothesis holds for H&R entrepreneurship:

Hypothesis 2: Being male is positively related to the propensity to create a firm in H&R sector.

*Household income*

According to Gollier (2002) and Guiso et al. (2002, 2003) one of the determinants of how much of their household income people invest in risky assets is their net wealth and income level. High income levels allow individuals to distribute their wealth in a wider range of investments, including riskier ones (Maula, et al., 2003). Investing part of the household income in creating a firm is a risky investment, so this work will test the following hypothesis:

Hypothesis 3: The higher the household income, the higher the propensity to create a firm in H&R sector.

*Work status*

From a demographic perspective researchers have generally argued that people in work are more likely to create a firm than people not in work (Arenius and Minniti, 2005). Moreover, incubator theory (Cooper, 1985; Cooper and Bruno, 1977) holds that many business projects or ideas responsible for turning a wage worker into an entrepreneur start life in the working environment. What is not so clear is whether unemployment discourages firm creation by making it appear too difficult or, in contrast, encourages people to try their hand at entrepreneurship to raise their incomes (Blanchflower, 2004).

Nevertheless, adopting the perspective of marginalization theory, some empirical studies confirm that marginalization encourages firm creation. Thus Evans and Leighton (1989) find that unemployment and poor working conditions increase the probability that people will create their own employment.

In view of the opposing arguments from the literature, the following hypothesis aims to decide whether the relation—if there is one—between employment and firm creation is positive or not:

Hypothesis 4: Being in employment is positively related to the propensity to create a firm in H&R sector.

**Perceptual variables**

*Perception of opportunities*

According to the theory of planned behavior (Ajzen, 1991), individuals' attitude influences their behavior. This attitude is the result of the individuals' favorable or unfavorable evaluation of that behavior. When individuals ask themselves if there are any business opportunities, what they are really doing is evaluating their own confidence in the economic climate (Maula, et al., 2003). If the individuals' evaluation is positive, their attitude toward entrepreneurial behavior should be favorable.

Entrepreneurs' ability to spot business opportunities is influenced by the number of such opportunities in their environment. But as Kirzner's entrepreneurship theory recognizes, entrepreneurs have a special "alertness" that allows them to spot opportunities that others miss (Kirzner, 1979). Consequently, faced by a similar number of opportunities in a particular environment, an alert entrepreneur perceives more opportunities than an ordinary person does. This idea is tested in the following hypothesis:

Hypothesis 5: The greater the number of business opportunities perceived, the higher the propensity to create a firm in H&R sector.

*Fear of failure*

The theory of planned behavior (Ajzen, 1991) holds that individuals' fear of failure leads to the perception that they are unable to control the behavior required to create a firm. This generates an unfavorable attitude toward such behavior. The absence of this fear would eliminate the perception of inability to control the situation and hence the unfavorable attitude toward such behavior.

On the other hand, fear of failure is closely related to risk aversion. The greater the individuals' risk aversion, the more they will fear failure. Various empirical studies using the trait theory perspective (McClelland, 1961; Collins and Moore, 1964) show that entrepreneurs prefer to take moderate risks (Amit et al., 1993; Brockhaus, 1976; Mancuso, 1975; McClelland and Winter, 1970). However, moderate risk aversion sometimes results from entrepreneurs perceiving that their chances of failure are low, and so their perception of the existing risk is also low (Amit et al., 1993). This does not mean that the real risk will be lower, but that the perceived risk is lower, and consequently the entrepreneur's fear of failure too. The following hypothesis captures the above ideas:

Hypothesis 6: The lower the fear of failure, the higher the propensity to create a firm in H&R sector.

*Perception of social legitimacy*

The theory of planned behavior (Ajzen, 1991) considers that a favorable or unfavorable attitude toward a particular behavior—in this case, firm creation—is influenced by social norms that either encourage or discourage this behavior. In this context, individuals' perceptions of what their social environment considers acceptable or not will influence their propensity to create a firm. Such factors include: whether the individual perceives that people regard being an entrepreneur as an attractive profession; and whether the individual perceives that people think successful entrepreneurs have a high social status and prestige. The next hypotheses follow on from this:

Hypothesis 7: The perception that society regards being an entrepreneur as an attractive profession is positively related to the propensity to create a firm in H&R sector.

Hypothesis 8: The perception that society believes that successful entrepreneurs gain high social status and prestige is positively related to the propensity to create a firm in H&R sector.

**Intellectual and social capital**

According to Davidsson and Honig (2003), intellectual and social capital are important factors in the decision to create a firm and in its subsequent success or failure. The theory of intellectual capital considers that knowledge improves individuals' cognitive skills and allows them to undertake activities more productively and efficiently (Schultz, 1959; Becker, 1964; Mincer, 1974). Individuals with a higher-quality intellectual capital should also be better at detecting the existence of profitable business opportunities (Davidsson and Honig, 2003).

The theory of social capital is closely related to social network theory. Both theories consider individuals' ability to extract benefits from the members of their social network (Lin et al., 1981; Portes, 1998). Social capital offers individuals social exchanges (Emerson, 1972) and the exchange of resources and information that will be useful for creating a firm (Davidsson and Honig, 2003).

This section considers entrepreneurial ability as part of the individual's intellectual capital, and knowing other entrepreneurs as part of their social capital.

*Entrepreneurial ability: Confidence in one's skills*

The section on perceptual variables discusses fear of failure, and suggests that entrepreneurs have different perceptions about the risk of embarking on a business venture than the rest of the population. According to Amit et al. (1993), this perception of the risk is moderated by the confidence that this type of person has in their skills and abilities. The entrepreneur is able to handle high-risk situations, perceiving that the risk is lower due to their confidence in their ability to handle it.

Szivas (2001) considers the role of acquiring skills and knowledge through experience and suggests that in the tourism industry skills are relatively easy to acquire, which encourages entry. Thus this author argues that previous skills and knowledge are not so important in this industry.

In contrast, according to the theory of planned behavior (Ajzen, 1991), if individuals feel that they have the ability, knowledge and skills required to create a firm, they will perceive that they are in control of the entrepreneurial behavior, and they will probably have a positive attitude toward such behavior. According to this theory, the more the individuals believe that they can achieve a valuable objective (in this case, create a firm), the more likely they will behave in such a way as to achieve that objective. The following hypothesis captures this idea:

Hypothesis 9: Confidence in one's skills is positively related to the propensity to create a firm in H&R sector.

*Entrepreneurial ability: Educational level*

Some researchers treat educational level as a demographic variable (Arenius and Minnti, 2005), while others treat it as part of the entrepreneur's intellectual capital (Davidsson and Honig, 2003). The current work opts to include this variable in this latter category.

The abovementioned confidence in one's skills is not necessarily related to the educational level. In fact some authors find that entrepreneurs frequently possess a wide range of skills but not an advanced or specific education (Murphy et al., 1991; Leazar, 2002). This contradicts Blanchflower's (2004) finding that a high educational level is positively related to the creation of technology firms in rich countries. Arenius and Minniti (2005) believe that the problem is even more complex, since countries have different, non-equivalent educational systems and the term "entrepreneur" is very broad. Entrepreneurs that create hi-tech companies conceivably need a high educational level. On the other hand, if the entrepreneurs only exploit a market opportunity they have spotted, a high educational level will not always be necessary.

According to the theory of planned behavior (Ajzen, 1991) a high educational level is likely to be positively associated with a greater perception of control (Maula, et al., 2003). This education will consequently contribute to individuals' belief that they have sufficient ability to start up a firm successfully. The next hypothesis is as follows:

Hypothesis 10: The educational level is positively related to the propensity to create a firm in H&R sector.

*Knowing other entrepreneurs*

Various empirical studies stress the important positive effect of indirect experience on the propensity to create a firm (Delmar and Gunnarsson, 2000; Scherer et al., 1991). Adopting the perspective of the theory of planned behavior (Ajzen, 1991), personally knowing other entrepreneurs should generate positive attitudes toward entrepreneurs in general, breaking down mental barriers (Maula, et al., 2003). Thus knowing other entrepreneurs conceivably influences individuals' subjective norms with regard to firm creation. This would make firm creation if not exactly a desired behavior at least an accepted behavior. Knowing other entrepreneurs also improves individuals' perception that they are able to control the necessary actions for creating a firm. At least this perception of control would be greater than in the case of not knowing any entrepreneurs.

Moreover, considering role theory (Veciana, 2007) individuals who know other entrepreneurs either from their close geographic environment or from more or less direct relationships (friends, relatives, etc.) may hear about facts that make the possibility of creating a firm and being successful in the attempt seem credible. Thus individuals who can capture and replicate "entrepreneurial roles" will be more likely to become entrepreneurs too.

Finally, from the network theory perspective, social networks can provide entrepreneurs with key information, ideas and resources to launch their new firm (Larson and Starr, 1993). If some members of the H&R entrepreneur's social network are entrepreneurs, the information, ideas and resources will undoubtedly be of a higher quality. Contacts with entrepreneurs will also provide access to other entrepreneurs of interest to the new firm, and also guide the H&R entrepreneurs in their relationships with public authorities and financial institutions. The above reasoning leads to the penultimate hypothesis:

Hypothesis 11: Knowing other entrepreneurs increases the propensity to create a firm in H&R sector.

*Being a business angel*

On the other hand, business angels tend to make their investments with some previous knowledge of entrepreneurship (Maula, et al., 2003). Also, from the planned behavior perspective they are likely to have a favorable attitude toward entrepreneurial behavior, and this will have a consequent impact on their propensity to engage in such behavior (Ajzen, 1991). They are likely to have a quite moderate, rather than high, risk aversion. Thus they could consider that starting up a new firm is acceptable (Amit et al., 1993; Brockhaus, 1976; Mancuso, 1975; McClelland and Winter, 1970). Moreover, adopting role theory, business angels move among entrepreneurs and so they are likely to hear about credible success stories that make firm creation seem feasible for them personally (Veciana, 2007). Finally network theory suggests that like in the previous hypothesis, contact with other entrepreneurs (this time, through their previous role as a business angel) provides access to ideas, information and resources that are critical for creating and consolidating a firm (Larson and Starr, 1993). All this is tested in the final hypothesis:

Hypothesis 12: Being a business angel increases the propensity to create a firm in H&R sector.

**RESEARCH METHODOLOGY**

**Data**

The data used in this paper come from the Global Entrepreneurship Monitor (GEM) project. The authors tested the above hypotheses using a sample of 121,218 interviews with adults (18-64 years old) collected during spring 2008 in countries participating in the GEM project. Details about the procedures used to collect and harmonize GEM data can be found in Reynolds et al. (2005). This survey was carried out by companies experienced in market research and public opinion using a questionnaire designed to analyze the behavior of new entrepreneurs. Because of individual-level missing data, only 33,711 respondents were included in the final sample.

**Figure 1. Classifying entrepreneurially active respondents**

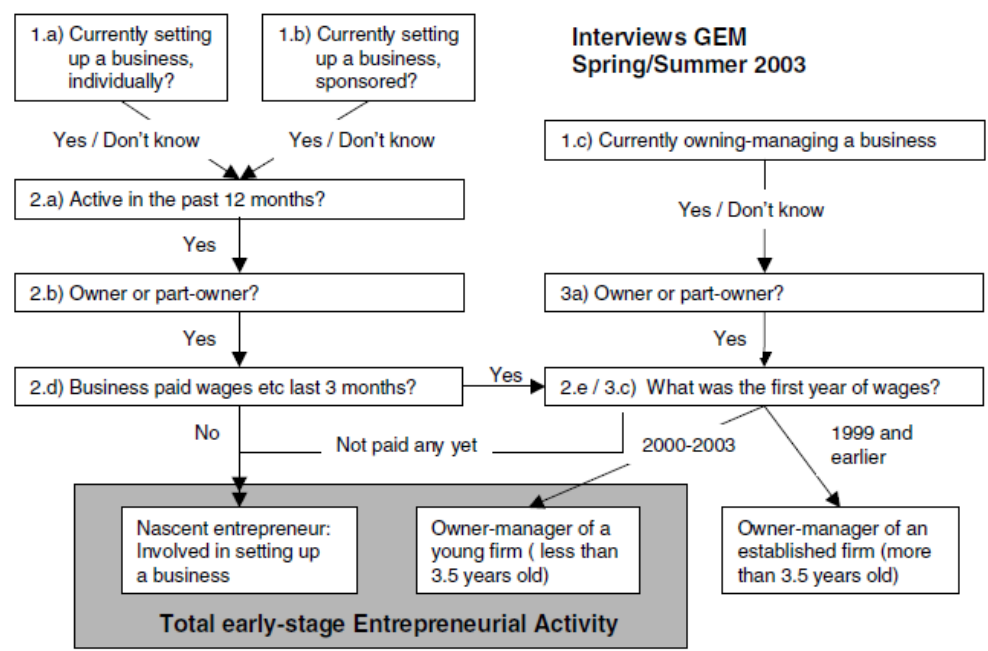

Source: Reynolds et al. (2005).

**Measures**

*Dependent variable*

The dependent variable is the early-stage H&R entrepreneur. Following Reynolds et al.'s (2005) criterion (see Figure 1), those entrepreneurs who have not paid salaries or wages for more than 3.5 years are considered to be involved in the early stages of the entrepreneurial process.
15

In all cases business types were recorded and categorized according to the UN Statistical Office classification[1].

The early-stage H&R entrepreneur was measured using a dichotomous variable taking value 1 when the respondents are in early-stage entrepreneurship and belong to division 55, "Hotels and restaurants", and 0 otherwise.

The relative frequency of early-stage H&R entrepreneurs in the current sample is 0.7%.

*Independent variables*

Demographic and economic factors

- Age: measured in years
- Gender: dichotomous variable taking value 0 for males and 1 for females
- Household income: the response categories were "Lowest 33 percentile", "Middle 33 percentile" and "Upper 33 percentile". In the logistic regression analysis, the first income group is used as the reference category
- Work status: the response categories were "full: full or part time", "part time only", "retired, disabled", "homemaker", "student", "not working, other"**.** In the logistic regression analysis, the last income group is used as the reference category.

Perceptual variables

- Opportunity perception: measured using a dichotomous variable taking value 1 if the respondent responds affirmatively to the question "In the next six months there will be good opportunities to start up new businesses in the area where you live", and 0 otherwise

---

[1] Department of Economic and Social Affairs (1990) International Standard Industrial Classification of All Economic Activities: Revision 3.

- Fear of failure: dichotomous variable equal to 1 if the respondent responds affirmatively to the question "Fear of failure would be a brake for you if you had to launch a business", and 0 otherwise.

Perception of social legitimacy

- Attractive profession: dichotomous variable taking value 1 if the respondent responds affirmatively to the question "In your country (region), most people believe that starting up a business is an attractive profession", and 0 otherwise
- Social status: dichotomous variable equal to 1 if the respondent responds affirmatively to the question "In your country (region), a person who successfully starts up a new business gains high social status and prestige", and 0 otherwise.

Intellectual and social capital

Entrepreneurial ability

- Confidence in one's skills: dichotomous variable equal to 1 if the respondent responds affirmatively to the question "You have the necessary knowledge, skills and experience to start up a new business", and 0 otherwise
- Educational level: the respondents were asked to state the highest level of education they had attained. Their responses were harmonized in all the participating countries into a five-category variable. The five categories are "none", "some secondary", "secondary degree", "post-secondary", and "university bachelor's degree or higher". In the logistic regression analysis, the first group is used as the reference category.

Social capital

- Knowing other entrepreneurs: dichotomous variable equal to 1 if the respondent responds affirmatively to the question "You know personally someone who has started up a new business in the past two years", and 0 otherwise
- Business angel: dichotomous variable equal to 1 if the respondent responds affirmatively to the question "In the past three years you have provided personal funds to help other people start up a business. Do not include investment in bonds, shares or mutual funds", and 0 otherwise.

**Econometric modeling**

*Specification*: given the nature of the research question posed—evaluating the influence of a series of independent variables on a dichotomous dependent variable—the appropriate econometric model is the general logistic regression model.

*Estimation*: the vector of unknown parameters of the model was estimated from the sample information using the maximum likelihood function as loss function. In order to evaluate the possible existence of multicollinearity between the model variables, the authors inspected the values of the matrix of sample correlations between the independent variables.

*Verification*: the authors verified the model using the likelihood ratio test, which verifies the statistical significance of all the model coefficients.

*Exploitation*: given that the objective is more for estimation than for prediction the exploitation of the estimated model focuses more on verifying the significance of the parameters and identifying factors of "risk" than on obtaining predictions of the dependent variable.

**RESULTS AND DISCUSSION**

Table I shows the descriptive statistics and the correlation matrix of the variables analyzed. As can be seen, the correlations between the exogenous variables are not high, which means that multicollinearity between the variables, and consequently the presence of undesired effects on the estimations of the parameters, are unlikely. Although, with large sample sizes, the significance of correlation coefficients may be misleading because the sampling distribution has less dispersion, and therefore smaller observed correlation coeficients (i.e. closer to 0) can reach the critical region and thus be statistically significant (Rubin, 2010).

In this work missing cases are those lacking a response to one or more questions. Some 72.2% of the total cases analyzed are missing (see Table II).

**Table II. Missing cases**

| Nonweighted cases(a) | | N | % |
|---|---|---|---|
| Cases selected | Included in analysis | 33711 | 27.8 |
|  | Missing cases | 87507 | 72.2 |
|  | Total | 121218 | 100.0 |
| Cases not selected | | 0 | .0 |
| Total | | 121218 | 100.0 |

In the adjusted model, shown in Table V, the authors entered all the variables. The omnibus test tests the null hypothesis that all the model coefficients are equal to zero, compared to the hypothesis that at least one parameter is nonzero. The null hypothesis can be rejected at the 1% level and hence the goodness of fit of the model is acceptable (see Table III).

**Table III. Omnibus test on model coefficients**

| | | Chi-square | df | Sig. |
|---|---|---|---|---|
| Step 1 | Step | 383.042 | 20 | .000 |
|  | Block | 383.042 | 20 | .000 |
|  | Model | 383.042 | 20 | .000 |

**Table I. Descriptive statistics and correlation matrix**

| | N | Mean | S.D. | 1 | 2 | 3 | 4 | 5 | 6 | 7 | 8 | 9 | 10 | 11 | 12 |
|---|---|---|---|---|---|---|---|---|---|---|---|---|---|---|---|
| 1. H&R entrepreneur | 121218 | .0070 | .08336 | | | | | | | | | | | | |
| 2. Age | 121218 | 39.36 | 13.005 | -.022(**) | | | | | | | | | | | |
| 3. Gender | 121218 | .50 | .500 | -.007(*) | .016(**) | | | | | | | | | | |
| 4. Household income | 96186 | 1.95 | .790 | .016(**) | -.031(**) | -.086(**) | | | | | | | | | |
| 5. Perception of opportunities | 77444 | .37 | .482 | .039(**) | -.078(**) | -.062(**) | .070(**) | | | | | | | | |
| 6. Fear of failure | 91668 | .40 | .490 | -.025(**) | .022(**) | .064(**) | -.058(**) | -.083(**) | | | | | | | |
| 7. Attractive profession | 82484 | .67 | .469 | .015(**) | -.041(**) | -.003 | -.024(**) | .108(**) | .006 | | | | | | |
| 8. Social status | 85194 | .68 | .468 | -.001 | -.036(**) | -.009(**) | -.011(**) | .114(**) | .017(**) | .181(**) | | | | | |
| 9. Perception of entrepreneurial ability | 91302 | .54 | .498 | .069(**) | -.018(**) | -.125(**) | .091(**) | .192(**) | -.162(**) | .029(**) | .042(**) | | | | |
| 10. Educational level | 121218 | 2.29 | 1.186 | .000 | -.088(**) | -.023(**) | .222(**) | -.016(**) | -.012(**) | -.077(**) | -.018(**) | .056(**) | | | |
| 11. Knowing other entrepreneurs | 94103 | .44 | .496 | .045(**) | -.104(**) | -.094(**) | .105(**) | .201(**) | -.050(**) | .020(**) | .039(**) | .242(**) | .084(**) | | |
| 12. Business angel | 120803 | .04 | .192 | .034(**) | -.011(**) | -.043(**) | .053(**) | .084(**) | -.029(**) | .006 | .011(**) | .107(**) | .033(**) | .147(**) | |
| 13. Work status | 114723 | 2.23 | 1.705 | -.028(**) | -.171(**) | .192(**) | -.156(**) | -.002 | .005 | .034(**) | .024(**) | -.113(**) | -.166(**) | -.059(**) | -.031(**) |

\*\* Correlation significant at 0.01 level (2-tailed)
\* Correlation significant at 0.05 level (2-tailed)

The authors also used Hosmer and Lemeshow's goodness-of-fit test, which basically measures the extent to which the predicted and observed probabilities coincide, so that if the fit is good, a high value in the predicted probability will be associated with the result Y=1 in the response variable. The difference in frequencies then distributes as a chi square and can be tested statistically. The hypothesis of an adequate model fit is accepted if p>0.05. In this case, as Table IV shows, the model has an acceptable fit.

**Table IV. Hosmer and Lemeshow test**

| Step | Chi-square | df | Sig. |
|---|---|---|---|
| 1 | 12.304 | 8 | .138 |

Table V shows the results of the logistic regression including all the variables available for the complete sample.

**Table V. Results of logistic regression (dependent variable: H&R entrepreneur)**

|  | B | SE | Wald | df | Sig. | Exp(B) |
|---|---|---|---|---|---|---|
| **Demographic and economic** | | | | | | |
| Age | -.029 | .004 | 44.046 | 1 | .000 | .971 |
| Gender | .410 | .095 | 18.437 | 1 | .000 | 1.507 |
| Household income | | | 23.082 | 2 | .000 | |
| -Middle 33%tile | .592 | .123 | 23.078 | 1 | .000 | 1.807 |
| -Upper 33%tile | .417 | .132 | 9.912 | 1 | .002 | 1.517 |
| Work status | | | 23.245 | 5 | .000 | |
| -Full: full or part time | .438 | .226 | 3.770 | 1 | .052 | 1.549 |
| -Part time only | .462 | .266 | 3.022 | 1 | .082 | 1.587 |
| -Retired, disabled | .244 | .410 | .354 | 1 | .552 | 1.276 |
| -Homemaker | .018 | .299 | .003 | 1 | .953 | 1.018 |
| -Student | -.852 | .367 | 5.394 | 1 | .020 | .427 |
| **Perceptual** | | | | | | |
| Opportunities | .266 | .095 | 7.859 | 1 | .005 | 1.304 |
| Fear of failure | -.340 | .107 | 10.185 | 1 | .001 | .712 |
| Attractive profession | .140 | .105 | 1.782 | 1 | .182 | 1.151 |
| Social status | -.119 | .099 | 1.454 | 1 | .228 | .888 |
| **Entrepreneurial ability** | | | | | | |
| Confidence in one's skills | 1.343 | .161 | 69.585 | 1 | .000 | 3.832 |
| Educational level | | | 10.000 | 4 | .040 | |
| -some secondary | .210 | .426 | .242 | 1 | .623 | 1.233 |
| -secondary degree | .017 | .426 | .002 | 1 | .968 | 1.017 |
| -post-secondary | .054 | .430 | .016 | 1 | .899 | 1.056 |
| -bachelor's degree or higher | -.213 | .428 | .247 | 1 | .619 | .808 |
| **Social capital** | | | | | | |
| Knowing other entrepreneurs | .292 | .100 | 8.488 | 1 | .004 | 1.339 |
| Business angel | .314 | .143 | 4.851 | 1 | .028 | 1.369 |
| Constant | -5.583 | .530 | 110.841 | 1 | .000 | .004 |

\*\*\* significant at p≤0.001; \*\* significant at p≤0.01; \* significant at p≤0.05

Among the sociodemographic variables three have coefficients significantly different from zero: age, gender and household income. Age's coefficient has a negative sign (-0.029), indicating that the probability of being an H&R entrepreneur declines with age. This result confirms Hypothesis 1, and is coherent with the results of other studies examining this relation for other types of entrepreneur (Lévesque and Minniti, 2006; Arenius and Minniti, 2005).

The regression coefficient of gender has a positive sign. This suggests that the probability of becoming a hospitality entrepreneur is greater for women, which contradicts the postulate in Hypothesis 2. Moreover, the odds ratio of this variable is 1.507, which indicates that women are 50% more likely to be an H&R entrepreneur than men. This result also contradicts the results of other studies examining this relation for nascent entrepreneurship (Reynolds et al., 2003; Arenius and Minniti, 2005), and suggests that the variable gender behaves differently in this sector.

Hypothesis 3 postulates a positive relation between household income and propensity to be an H&R entrepreneur, in other words the propensity increases with income level. The results obtained here confirm this relation, because belonging to the middle income category almost doubles the probability of being a hospitality entrepreneur with respect to the lowest income level (odds ratio=1.807). The probability also increases for the upper income category: individuals from this level are 50% more likely to start a business in this sector than those from the reference category. This result is coherent with earlier research (Maula, et al., 2003).

In the current model work status does not show a clear statistically significant influence. The coefficient of this variable is significantly different from zero for the variable measured as a whole. But only one category shows a clear negative relation with the probability of being an entrepreneur in the hospitality industry: the status of student. In other words, the group of students is less likely to start a business in this sector than the reference category (not working). The odds ratio of this variable is 0.427, indicating that students are half as likely to start a business in this sector as people not working.

Only the first two perceptual variables are significant. In particular, perception of opportunities is positively related to being an H&R entrepreneur. This result coincides with those of other studies examining this relation for other entrepreneurs, and could provide support for Kirzner's theory, according to which "alertness" is a necessary condition in the process of opportunity recognition.

In turn, fear of failure has a negative impact on being an H&R entrepreneur. This means that individuals who fear failure are less likely to start a business in this sector. This result is consistent with findings for other types of entrepreneur (Arenius and Minniti, 2005; Weber and Milliman, 1997). These latter authors find that a high perception of risk reduces individuals' incentives to start a business.

The coefficients of the two perceived social legitimacy variables in the model—attractive profession and social status—are nonsignificant, so the results do not support the existence of a significant relation between each variable and the probability of being an entrepreneur in the H&R sector.

In contrast, the individual's perception of having the skills and ability to start a business has a significant, positive relation with the probability of being an H&R entrepreneur, which confirms Hypothesis 9. In fact, according to the results this is the variable analyzed that has the strongest effect on the dependent variable, since its odds ratio is 3.832. Thus individuals who feel capable of starting a business are almost four times more likely to do so than the rest of the adult population. This result for entrepreneurs in the tourist sector coincides with the results of work examining other types of entrepreneur, which find that confidence in one's own ability to start a business is the most influential component in the decision to start a business (Koellinger et al., 2004; Arenius and Minniti, 2005).

The other entrepreneurial ability variable analyzed here—educational level—gives nonsignificant values in all categories compared to the reference category (none, i.e., no education). Thus the results do not provide support for Hypothesis 10. This finding may be indicating that educational level behaves differently in the H&R sector, since other studies consulted show a positive relation between these two variables, meaning that a high education level increases intellectual capital, and consequently the ability to recognize and exploit business opportunities.

Finally, the two social capital variables in the model—knowing other entrepreneurs and financing or having financed another business as a business angel—both have a significant, positive effect. These results provide support for hypotheses 11 and 12, and are consistent with other studies involving other types of entrepreneur (Arenius and Minniti, 2005; Minniti 2004). That knowing other entrepreneurs can have a positive effect on the decision to start a business may be because belonging to social networks reduces ambiguity and uncertainty, facilitates the exchange of information (Weber and Milliman, 1997), resources and personal contacts, and shows that taking on the entrepreneur role is plausible (Veciana, 2007; Kent et al., 1981; Cooper and Dunkelberg, 1987; Duchesnau and Gartner, 1990).

**CONCLUSIONS**

Entrepreneurship research is a relatively young and emerging discipline. This is particularly true in studies in the tourism sector, in which few researchers have studied the firm creation phenomenon at depth (Ateljevic and Page, 2009).

This work represents a first step in understanding the factors that influence the decision to start a business in the tourism subsector of Hotels and Restaurants (H&R) and a new scientific contribution based on the use of GEM data.

The authors have analyzed and assessed the influence on the propensity to be an H&R entrepreneur of a series of demographic and economic factors, variables related to perceptions of the environment and personal traits, and variables measuring intellectual and social capital. For this purpose, they used the Adult Population Survey, a large sample that is a source of empirical evidence for the Global Entrepreneurship Monitor (GEM) project, of which the current authors are members.

The statistical analysis used logistic regression techniques appropriate for analyzing this type of research question. They were based on the identification of the influence of certain factors on a dichotomous dependent variable.

The findings suggest that the three variables age, gender, and household income affect the decision to become an entrepreneur in H&R segments. Specifically, individuals are more likely to be an H&R entrepreneur when they are young and female and their household income is in the middle or upper categories.

Besides the abovementioned demographic and economic factors, other factors measuring the individuals' perception of their environment and their personality traits affect that decision. The inclusion of these perceptual variables in the model used in this work is highly important because authors do not usually include them in their analytical models of entrepreneurial behavior (Arenius and Minniti, 2005). The results obtained here show that individuals who perceive good business opportunities in their close environment and who do not fear failure are more likely to start a business in the H&R sector.

Another interesting result of this work is that it cannot offer empirical support for the idea that perceiving the profession of entrepreneur to be attractive and to have social prestige affects the decision to become an entrepreneur.

On the other hand, according to this model, the most important factor behind the decision to start a business in the H&R sector is having confidence in one's skills, knowledge and experience in startups. In contrast, the current results do not empirically confirm that higher educational level affects the decision to start a business in the hospitality industry. A clear practical implication for policymakers in the areas of training, employment orientation and firm creation in the sector derives from these two results. They should concentrate their efforts on helping the population to develop the skills needed to start a business. Advanced academic knowledge is less important than practical and professional knowledge to start an H&R business.

Finally, the two variables measuring social capital—knowing other entrepreneurs and having invested in another business as a business angel—seem to have a critical influence on the decision to start a business in the H&R sector. Again, practical implications derive from this result. Policymakers should encourage and facilitate relationships between individuals, since this will help reduce uncertainty and boost the exchange of that valuable resource, information.

In future research the current authors aim to examine the role of intercountry differences in more detail. For this purpose, the countries could be grouped into three stages of economic development as defined by the World Economic Forum's Global Competitiveness Report[2]: factor-driven, efficiency-driven and innovation-driven (Porter, et al., 2002; Bosma and Levie, 2010).

Finally, it would be particularly interesting to investigate H&R entrepreneurship in the first stage of the entrepreneurship process—the opportunity recognition stage. The current authors aim to compare the ability to recognize business opportunities between H&R entrepreneurs and the rest of the entrepreneurs (Ramos-Rodriguez et al., 2011). This analysis would shed more light on the entrepreneurship process in this sector.

---

[2] This report is available in www.gemconsortium.org